\documentclass[aps,12pt,final,notitlepage,oneside,onecolumn,nobibnotes,nofootinbib,superscriptaddress,showpacs,centertags]{revtex4}

\usepackage{graphicx}
\usepackage{bm}
\begin{document}

\title{The one-particle momentum transfer in point-form spectator approximation}
\author{N.A. Khokhlov}
\email{khokhlov@fizika.khstu.ru} \affiliation{\it Pacific State
University, 680035, Khabarovsk, Russia}
\date{\today}
\begin{abstract}
The one-particle momentum transfer operator  is derived in
point-form spectator approximation for $NN$ system. General
expression is applied to elastic electron-deuteron scattering and to
deuteron photodisintegration.
\end{abstract}
\maketitle

\section{\label{sec:intro}INTRODUCTION}
In Ref.~\cite{AllenFF} the elastic electron-deuteron scattering was
described in frames of point-form (PF) of relativistic quantum
mechanics (RKM). It was shown that in PF spectator approximation
 that the
momentum of the unstruck particle (the spectator) is unchanged,
while the impulse given to the struck particle is not the impulse
given to the deuteron. In present paper this result is generalized
and one-particle momentum transfer operator is derived for arbitrary
$NN$-system following a general approach to construction of the
electromagnetic current operator for relativistic composite system
\cite{Lev}.

We define the momentum transfer $Q^2_i$ to the $i$-particle as an
increment of $i$-particle 4-momentum $q_i$ \cite{AllenFF}%
\begin{equation}
Q^2_i=|(q_i'-q_i)^2|.
\end{equation}
For interacting particles the individual 4-momenta are not defined
before photon absorption (emission) as well as after it.
Therefore we introduce operator ${ Q}^2_i$ corresponding to the
physical quantity of  the momentum transfer ${Q}^2_i$.

The plan of the paper is as follows. In Sect. 2 we derive the
one-particle momentum transfer operator. In Sect. 3 we show that our
result is equivalent to the corresponding result of \cite{AllenFF}
for elastic electron-deuteron scattering. In Sect. 4 we consider
deuteron photodisintegration.

\section{\label{sec:operator} The one-particle momentum transfer operator}

In the general case there are initial $NN$-state and associated
initial s.c.m. (i.s.c.m.), and final $NN$-state  and associated
final s.c.m. (f.s.c.m.) Suppose that the photon momentum (momentum
transfer) is along the $z$ axis. Values of photon momentum and
energy in i.s.c.m. are $|{\bf q}_{\gamma}|$ and $q_{\gamma}^0$
correspondingly. Momentum transfer is $Q^2=|{\bf q}_{\gamma}|^2-
(q_{\gamma}^0)^2$. Let $P$ be the total 4-momentum of the
$NN$-system, $M$ be the mass of the $NN$-system, $G=P/M$ be the
system 4-velocity. Index $i(f)$ means initial (final) state of the
$NN$-system. Transformation from i.s.c.m. to the special system
suggested by Lev \cite{Lev} (L.s.) where
\begin{equation}
{\bf G}_f+{\bf G}_i=0|_{L.s.}\label{Levscm}%
 \end{equation}
is defined by angle $\Delta/2$ such that
\begin{equation}
\tanh\Delta/2=h, \label{Deltafromh}
\end{equation}
where ${\bf h}={\bf G}_f/G^{0}_f|_{L.s.}$. The Lev frame
(\ref{Levscm}) is not equivalent to the Breit frame  defined by the
condition ${\bf P}_f+{\bf P}_i=0$ if $M_f\neq M_f$. In case of
elastic electron-deuteron scattering these frames coincide.

From this point we may use a special derivation of Ref.
\cite{AllenFF} (Eqs. (\ref{initialmomenta}-\ref{transferMy} of the
present paper)).

The initial energies and z-components of momenta in L.s. are
\begin{eqnarray}
w_1=w\cosh \Delta/2+q_{z}\sinh\Delta/2\nonumber\\
q_{1z}=q_{z}\cosh \Delta/2+w\sinh\Delta/2\nonumber\\
w_2=w\cosh \Delta/2-q_{z}\sinh\Delta/2\nonumber\\
q_{2z}=-q\cosh \Delta/2+w\sinh\Delta/2,\label{initialmomenta}
\end{eqnarray}
where ${\bf q}$ and $w=\sqrt{{\bf q}^2+m^2}$ are center of momentum
variables, ${\bf q}$ is momentum of particle one (internal variable)
After the photon absorption the $z$-component of the internal
variable and corresponding energy change
\begin{eqnarray}
q_{z}'=q_{z}\cosh \Delta\mp w\sinh\Delta\\
w'=w\cosh \Delta\mp q_{z}\sinh\Delta,
\end{eqnarray}
where the minus (plus) sign is used when particle one (two) is
struck. The final energies and momenta in L.s. will then be
\begin{eqnarray}
w_1'=w\cosh 3\Delta/2-q_{z}\sinh 3\Delta/2\nonumber\\
q_{1z}'=q_{z}3\cosh \Delta/2-w\sinh3\Delta/2\nonumber\\
w_2'=w_2;\ \ \ \ \ q_{2z}'=q_{2z}, \label{new_moment}
\end{eqnarray}
other components do not change. Some hyperbolic trigonometry reveals
that
\begin{equation}
(q_1'-q_1)^2=4(q_{z}^2-w^2)\sinh^2\Delta\label{transferMy},%
 \end{equation}
it follows from Eq.~(\ref{Deltafromh}) that
\begin{eqnarray}
\sinh\Delta=\frac{2h}{1-h^2}.%
 \end{eqnarray}
Since
\begin{equation}
q_{z}^2-w^2=-(m^2+{\bf q}^2_{\perp})=-(m^2+{\bf q}^2-\frac{({\bf q}\cdot{\bf h})^2}{h^2}),\label{q_zaction}%
 \end{equation}
 the momentum transfered to the struck particle is
\begin{equation}
Q^2_1=-(q_1'-q_1)^2=16(m^2+{\bf q}^2-\frac{({\bf q}\cdot{\bf h})^2}{h^2})\frac{h^2}{(1-h^2)^2}. \label{fatransfer}%
 \end{equation}
This is the general expression of the  ${ Q}^2_1={ Q}^2_2$, in case
of free two-particle states (for particles of equal masses). The
parameter ${\bf h}$ does not depend on interaction and is specified
by relative "position"\hspace{1mm} of i.s.c.m. and f.s.c.m.  In case
 of two interacting particles ${\bf q}$ and $Q^2_i$ are
operators in the internal space. In impulse representation ${\bf q}$
is a variable of integration \cite{AllenFF}. It is obvious that
forcing on the plane wave this operator is equivalent to the
multiplication by number ${ Q}^2_1={Q}^2_2>0$ if $h\neq 0$.

It may be convenient to express $\sinh\Delta$ in
Eq.~(\ref{transferMy}) through the invariant masses of initial
($M_i$) and final ($M_f$) $NN$-states. With transition from i.s.c.m.
to the special L.s. the 4-velocities of the initial and final states
are transformed as
\begin{eqnarray}
(1,0,0,0)|_{i.s.c.m.}\rightarrow (\frac{P^{0}_i}{M_i},0,0,-|{\bf G}_i|)|_{L.s.}\nonumber\\
(\frac{M_i+q_{\gamma}^{0}}{M_f},0,0,\frac{|{\bf
q}_{\gamma}|}{M_f})|_{i.s.c.m.}\rightarrow
(\frac{P^{0}_f}{M_f},0,0,|{\bf G}_f|)|_{L.s.},\label{toLevscm}
\end{eqnarray}
correspondingly, where order of components is
$(a^{0},a^{x},a^{y},a^{z})$. Lorentz transformations are linear,
therefore sum of these 4-velocities transforms as
\begin{eqnarray}
(\frac{M_f+M_i+q_{\gamma}^{0}}{M_f},0,0,\frac{|{\bf
q}_{\gamma}|}{M_f})|_{i.s.c.m.}\rightarrow
(\frac{P^{0}_f}{M_f}+\frac{P^{0}_i}{M_i},0,0,-|{\bf G}_i|+|{\bf G}_f|)|_{L.s.}\equiv\nonumber\\
\equiv(\frac{P^{0}_f}{M_f}+\frac{P^{0}_i}{M_i},0,0,0)|_{L.s.}.\label{toLevscmSUM}
\end{eqnarray}
$z$-Component of this sum in L.s. is
\begin{equation}
\frac{|{\bf q}_{\gamma}|}{M_f}\cosh\Delta/2-\frac{M_f+M_i+q_{\gamma}^{0}}{M_f}\sinh\Delta/2=0,%
 \end{equation}
 whence
\begin{equation}
\sinh\Delta/2=\sqrt{\frac{|{\bf q}_{\gamma}|^2}{(M_f+M_i+q_{\gamma}^{0})^2-|{\bf q}_{\gamma}|^2}},\label{MyDelta}%
 \end{equation}
 and
\begin{eqnarray}
\sinh^2\Delta=4\sinh^2\Delta/2\cosh^2\Delta/2=\nonumber\\
=4{\frac{|{\bf q}_{\gamma}|^2}{(M_f+M_i+q_{\gamma}^{0})^2-|{\bf
q}_{\gamma}|^2}}({1+\frac{|{\bf q}_{\gamma}|^2}
{(M_f+M_i+q_{\gamma}^{0})^2-|{\bf q}_{\gamma}|^2}})=\nonumber\\
=4{\frac{|{\bf
q}_{\gamma}|^2(M_f+M_i+q_{\gamma}^{0})^2}{((M_f+M_i+q_{\gamma}^{0})^2-|{\bf q}_{\gamma}|^2)^2}}.\label{sinhDelta}%
 \end{eqnarray}
\section{\label{sec:ed-ed} Elastic electron-deuteron scattering}
In case of elastic electron-deuteron scattering $M_i=M_f=m_D$ ($m_D$
is mass of deuteron),
$\frac{P^{0}_i}{M_i}=\frac{P^{0}_f}{M_f}|_{L.s.}$ therefore L.s.
becomes the Breit system. Transformations  (\ref{toLevscm}) give for time components%
\begin{eqnarray}
\frac{P^{0}_i}{M_i}|_{L.s.}=\cosh\Delta/2=\nonumber\\
=\frac{P^{0}_f}{M_f}|_{L.s.}=\frac{m_D+q_{\gamma}^{0}}{m_D}\cosh\Delta/2-\frac{|{\bf
q}_{\gamma}|}{m_D}\sinh\Delta/2,
\end{eqnarray}
and for $z$-components  of 4-velocities
\begin{eqnarray}
|{\bf G}_i|_{L.s.}=-\sinh\Delta/2=\nonumber\\
=-|{\bf G}_i|_{L.s.}=
-(-\frac{m_D+q_{\gamma}^{0}}{m_D}\sinh\Delta/2+\frac{|{\bf
q}_{\gamma}|}{m_D}\cosh\Delta/2),
\end{eqnarray}
where $q^0_{\gamma}$ and ${\bf q}_{\gamma}$ are energy and momentum
of virtual photon correspondingly in the i.s.c.m. In this case
i.s.c.m. coincides with laboratory system.   Therefore
\begin{eqnarray}
q_{\gamma}^0\cosh\Delta/2=|{\bf q}_{\gamma}|\sinh\Delta/2\nonumber\\
(2m_D+q_{\gamma}^0)\sinh\Delta/2=|{\bf
q}_{\gamma}|\cosh\Delta/2.\label{f1Allen}
\end{eqnarray}
Product of l.h.s.'s is equal to product of r.h.s.'s:
\begin{eqnarray}
q_{\gamma}^0(2m_D+q_{\gamma}^0)=|{\bf q}_{\gamma}|^2,
\end{eqnarray}
i.e.
\begin{equation}
Q^2=|{\bf q}_{\gamma}|^2-(q_{\gamma}^0)^2=2m_{D}q_{\gamma}^0.
\end{equation}
Exclusion of $|{\bf q}_{\gamma}|$ from Eqs. (\ref{f1Allen}) gives
\begin{equation}
\tanh^2\Delta/2 =\frac{q_{\gamma}^0}{2m_{D}+q_{\gamma}^0},
\end{equation}
Exclusion of $q_{\gamma}^0$ from last two gives
\begin{equation}
\tanh^2\Delta/2 =\frac{Q^2}{Q^2+4m_{D}},
\end{equation}
finally
\begin{equation}
\sinh^2\Delta/2=\frac{Q^2}{4m_D^2},\label{deltaAllen}
 \end{equation}
i.e. expression of  \cite{AllenFF} then deuteron absorbs virtual
photon elastically, since.
\begin{equation}
\sinh\Delta=2\sqrt{\frac{Q^2}{4m_D^2}}\sqrt{1+\frac{Q^2}{4m_D^2}}.\label{deltaAllen}
 \end{equation}
 Therefore expression inferred in Ref.
\cite{AllenFF} is a special case of Eq.~(\ref{fatransfer}).
\section{\label{sec:ed-ed} Photodisintegration of deuteron}
In case of deuteron photodisintegration the photon is real. In some
cases (neglecting the final state interaction, or in some
calculation schemes) we need to find action of operator
(\ref{fatransfer}) on free final state described by plain wave.
 For real photon $|{\bf q}_{\gamma}|=q_{\gamma}^0=E_{\gamma}$. From
 (\ref{sinhDelta}) it follows that
\begin{eqnarray}
\sinh^2\Delta=4{\frac{E_{\gamma}^2(M_f+m_D+E_{\gamma})^2}{((M_f+M_i+E_{\gamma})^2-E_{\gamma}^2)^2}},\label{OursinhDelta}%
 \end{eqnarray}
 where $M_i=m_D$,  $E_{\gamma}$ is photon energy  in the i.s.c.m. In this case i.s.c.m.
coincides with laboratory system.  Since
$M_f=\sqrt{(m_D+E_{\gamma})^2-E_{\gamma}^2}$ then
\begin{eqnarray}
4{\frac{(M_f+m_D+E_{\gamma})^2}{((M_f+M_i+E_{\gamma})^2-E_{\gamma}^2)^2}}=%
\frac{1}{(m_D+E_{\gamma})^2-E_{\gamma}^2},
 \end{eqnarray}
therefore
 \begin{eqnarray}
\sinh^2\Delta=\frac{E_{\gamma}^2}{(m_D+E_{\gamma})^2-E^2_{\gamma}},\label{OursinhDelta2}%
 \end{eqnarray}
this is a kinematical factor. Other factor in Eq.~(\ref{q_zaction})
contains operator ${\bf q}^2_{\perp}$. This operator acting on the
plane wave in final state gives square of the proton momentum
component in f.s.c.m.  This component is orthogonal to ${\bf h}$ and
therefore it does not change with transformations from  i.s.c.m. to
f.s.c.m. and to L.s. We find this quantity in i.s.c.m. using
momentum conservation
\begin{eqnarray}
E_{\gamma}^2=|{\bf q}_n+{\bf q}_p|^2=\nonumber\\
={\bf q}_n^2+{\bf q}_p^2+2(\pm\sqrt{w_n^2-m^2-{\bf
q}^2_{\perp}}\sqrt{w_p^2-m^2-{\bf q}^2_{\perp}}-{\bf q}^2_{\perp})=\nonumber\\
={\bf q}_n^2+{\bf q}_p^2+2(\pm\sqrt{w_n^2-x}\sqrt{w_p^2-x}-(x-m^2))
\end{eqnarray}
where $|{\bf q}_p|$ ($w_{p}$) and $|{\bf q}_n|$  ($w_{n}$) are
momenta (energies) of final proton and neutron in
 i.s.c.m. (laboratory system for $\gamma d\rightarrow pn$). We take into account that $({\bf q}_p)_{\perp}=-({\bf
q}_n)_{\perp}\equiv{\bf q}_{\perp}$. Independently on the relative
direction of the parallel components of proton and neuteron momenta
(i.e. on sign $\pm$) there is a single root
\begin{eqnarray}
x=-(q_z^2-w^2)=m^2+{\bf q}_{\perp}^2=\nonumber\\%
=\frac{({\bf q}_n^2+{\bf q}_p^2+2m^2+2w_n w_p^2-E_{\gamma}^2)({\bf
q}_n^2+{\bf q}_p^2+2m^2-2w_n
w_p^2-E_{\gamma}^2)}{4({\bf q}_n^2+{\bf q}_p^2+2m^2-w_n^2-w_p^2-E_{\gamma}^2)}=\nonumber\\%
=-\frac{((w_n+w_p)^2-E_{\gamma}^2)((w_n-w_p)^2-E_{\gamma}^2)}{4E_{\gamma}^2}=\nonumber\\%
=-\frac{((m_D+E_{\gamma})^2-E_{\gamma}^2)((w_n-w_p)^2-E_{\gamma}^2)}{4E_{\gamma}^2}.\label{xfound}
\end{eqnarray}
 Inserting
(\ref{OursinhDelta2}) and (\ref{xfound}) into (\ref{transferMy}) we
get action of the 4-momentum transfer operator in the internal space
in case of plane wave in final state
\begin{equation}
Q_1^2=-(q_1'-q_1)^2=E_{\gamma}^2-(w_n-w_p)^2=(2w_n-m_D)(2w_p-m_D).\label{finaltransferMy}%
 \end{equation}
Obviously $Q_2^2=Q_1^2$. The operator may be substituted by this
number only if the 3-vector parameter ${\bf h}$ is fixed and there
is a plane wave in the bra part of the matrix element.

\end{document}